\documentclass{pspum-l}

\usepackage{color}
\definecolor{niceblue}{rgb}{0,0,0.6}
\usepackage[a4paper,colorlinks,citecolor=niceblue,linkcolor=niceblue,urlcolor=niceblue,pdfpagemode=None]{hyperref}
\usepackage{latexsym}
\usepackage{color}
\usepackage[all,cmtip]{xy}

\usepackage{stmaryrd}
\usepackage{amsmath, amscd, amssymb, mathrsfs, accents, amsfonts}
\usepackage{url}
\usepackage[all]{xy}
\usepackage{longtable}
\usepackage{dsfont}
\usepackage{tikz}
\usetikzlibrary{decorations.pathmorphing}
\usetikzlibrary{calc}
\usetikzlibrary{decorations.markings}
\def\nicedashedcolourscheme{\shadedraw[top color=blue!22, bottom color=blue!22, draw=gray, dashed]}

\def\nicenotpalecolourscheme{\shadedraw[top color=blue!32, bottom color=blue!32, draw=white]}
\def\nicepalecolourscheme{\shadedraw[top color=blue!22, bottom color=blue!22, draw=white]}

\def\nicereallynocolourscheme{\shadedraw[top color=white!2, bottom color=white!25, draw=white]}
\def\nicedashedpalecolourscheme{\shadedraw[top color=blue!12, bottom color=blue!12, draw=gray, dashed]}
\definecolor{Myblue}{rgb}{0,0,0.6}

\newcommand{\I}{\text{i}}
\newcommand{\C}{\mathds{C}}

\newcommand{\Z}{\mathds{Z}}
\newcommand{\be}{\begin{equation}}
\newcommand{\ee}{\end{equation}}
\newcommand{\bes}{\begin{equation*}}
\newcommand{\ees}{\end{equation*}}

\newcommand{\Res}{\operatorname{Res}}

\newcommand{\str}{\operatorname{str}}

\newcommand{\Hom}{\operatorname{Hom}}
\newcommand{\End}{\operatorname{End}}

\newtheorem*{theorem}{Theorem}
\newtheorem*{proposition}{Proposition}
\numberwithin{equation}{section}

\def\eval{\operatorname{ev}}
\def\coev{\operatorname{coev}}

\usetikzlibrary{arrows,calc,decorations.pathreplacing,decorations.markings,shapes.geometric,shadows}
\tikzset{
    string/.style={draw=#1, postaction={decorate}, decoration={markings,mark=at position .51 with {\arrow[draw=#1]{>}}}},
    costring/.style={draw=#1, postaction={decorate}, decoration={markings,mark=at position .51 with {\arrow[draw=#1]{<}}}},
    ostring/.style={draw=#1, postaction={decorate}, decoration={markings,mark=at position .47 with {\arrow[draw=#1]{>}}}},
    ustring/.style={draw=#1, postaction={decorate}, decoration={markings,mark=at position .56 with {\arrow[draw=#1]{>}}}},
    oostring/.style={draw=#1, postaction={decorate}, decoration={markings,mark=at position .43 with {\arrow[draw=#1]{>}}}},
    uustring/.style={draw=#1, postaction={decorate}, decoration={markings,mark=at position .59 with {\arrow[draw=#1]{>}}}},
    directed/.style={string=blue!50!black}, 
    odirected/.style={ostring=blue!50!black}, 
    udirected/.style={ustring=blue!50!black}, 
    oodirected/.style={oostring=blue!50!black}, 
    uudirected/.style={uustring=blue!50!black},     
    redirected/.style={costring= blue!50!black},
}

\usetikzlibrary{fadings,decorations.pathreplacing}

\begin{document}

\title{\mbox{A toolkit for defect computations in Landau-Ginzburg models}}

\author{Nils Carqueville}
\address{Simons Center for Geometry and Physics}
\email{nils.carqueville@scgp.stonybrook.edu}

\author{Daniel Murfet}
\address{Mathematical Sciences Research Institute}
\email{dmurfet@msri.org}


\begin{abstract}
We review the results of~\cite{cm1208.1481} on orientation reversal and duality for defects in topological Landau-Ginzburg models, with the intention of providing an easily accessible toolkit for computations. As an example we include a proof of the main result on adjunctions in a special case, using Pauli matrices. We also explain how to compute arbitrary correlators of defect-decorated planar worldsheets, and briefly discuss the relation to generalised orbifolds. 
\end{abstract}

\maketitle

\section{Introduction}

There is a wide spectrum of reasons to study Landau-Ginzburg models, of which we name only but a few. A certain subclass of such models is believed to have infrared fixed points under renormalisation group flow that describe `stringy' string vacua in the moduli space of Calabi-Yau compactifications~\cite{gvw1989, w9301, hhp0803.2045}. On the one hand, this CY/LG correspondence makes Landau-Ginzburg models a central player in the world of (homological) mirror symmetry. On the other hand it embeds into the more general CFT/LG correspondence which states that many properties of the conformal fixed point can already be directly described on the level of Landau-Ginzburg models. By now much circumstantial evidence has been collected in the bulk, boundary and defect sectors, see e.\,g.~\cite{m1989, vw1989, howewest, add0401, bg0506208, err0508, cr0909.4381, cr1006.5609}. Out of this (together with an increasing understanding of matrix factorisations) emerges the hope of better control over non-rational conformal field theories, as Landau-Ginzburg models appear comparably indifferent towards additional symmetries. Finally, Landau-Ginzburg models also feature in the construction~\cite{kr0401268} of homological link invariants~\cite{gw0512298, khovhompaper, MR2reps}; the results reported in this note may help to establish the precise connection to the work of~\cite{gsv0412243, w1101.3216}. 

Regarding the subject of defects in two-dimensional (topological) field theories we refer e.\,g.~to~\cite{k1004.2307, dkr1107.0495} as excellent introductions. Generally speaking one should take the defect sector of a theory as seriously as (its special case) the boundary sector. The fact that the former's target space interpretation is currently much less clear than the latter's in terms of D-branes poses a challenge whose resolution may significantly advance our view on string theory. Among the numerous important applications of defects we single out their use for renormalisation group flow~\cite{br0712.0188, g1201.0767}, in deforming conformal field theories away from the rational point~\cite{r1004.1909, br1211.4726}, their role in the AGT correspondence~\cite{aggtv0909.0945}, as well as their relevance for the refined link invariant constructions mentioned above. 

\medskip

The main part of this note is organised as follows. In Section~\ref{sec:defectsinLGmodels} we start by recalling the description of defects in topological Landau-Ginzburg models in terms of matrix factorisations. Then we explain the origin and nature of the type of duality we wish to consider, to wit adjunctions between defects. Our main theorem states the existence of these adjunctions and provides explicit evaluation and coevaluation maps to satisfy the needs of the practically-minded. Throughout we use a rigorous diagrammatic language which lends itself to efficient computations. The fact that it also has a natural physical interpretation is the starting point for the applications discussed in Section~\ref{sec:apps}: any planar defect-decorated correlator can be straightforwardly computed from our results. To illustrate this we first derive a residue expression for the action of defects on bulk fields; among its special cases we recover the Kapustin-Li disc correlator and a formula for boundary states. Next we consider a genus-one worldsheet and give a simple one-line proof of the Cardy condition. We end with a brief look on the generalised orbifolds of~\cite{genorb}.

\section{Adjoint defects in Landau-Ginzburg models}\label{sec:defectsinLGmodels}

We recall from~\cite{kl0210, bhls0305, l0312, br0707.0922} that oriented \textsl{defects} $X: W\rightarrow V$ between topological Landau-Ginzburg models with potentials $W\in \C[x] \equiv \C[x_1,\ldots,x_n]$ and $V\in \C[z] \equiv \C[z_1,\ldots,z_m]$ are described by matrix factorisations of $V-W$. This means that $X = X^0\oplus X^1$ is a free $\Z_2$-graded $\C[z,x]$-module equipped with a twisted differential, i.\,e.~an odd operator~$d_X$ which squares to $(V-W) \cdot 1_X$. \textsl{Defect changing fields} $\Psi \in \Hom(X,Y)$ take values in the cohomology of the BRST operator $\Psi \mapsto d_Y\Psi - (-1)^{|\Psi|}\Psi d_X$. Throughout this note we stick to the above notation of variable numbers $n,m$, potentials $W,V$, and defect~$X$: 
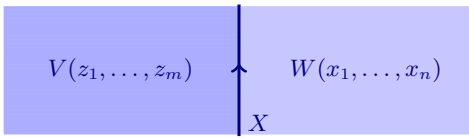
$$
\begin{tikzpicture}[very thick,scale=0.7,color=blue!50!black, baseline=0cm]
\nicepalecolourscheme (0,-1.25) rectangle (4.5,1.25);
\nicenotpalecolourscheme (-4.5,-1.25) rectangle (0,1.25);
\draw[line width=0] 
(0,1.25) node[line width=0pt] (A) {}
(0,-1.25) node[line width=0pt] (A2) {}; 
\draw[
	decoration={markings, mark=at position 0.55 with {\arrow{>}}}, postaction={decorate}
	]
 (0,-1.25) -- (0,1.25); 
 \draw[line width=0] 
(0,-1) node[line width=0pt, right] (Xbottom) {{\small $X$}}
(-3.8,0) node[line width=0pt, right] (Xbottom) {{\small $V(z_1,\ldots,z_m)$}}
(0.8,0) node[line width=0pt, right] (Xbottom) {{\small $W(x_1,\ldots,x_n)$}};
\end{tikzpicture}
$$

In the presence of topological defects there are two types of `multiplications': the \textsl{operator product} of fields given by matrix multiplication, and the \textsl{fusion} of defects and their fields. The latter is given by the tensor product $X \otimes_{\C[x]} Z$ with twisted differential $d_X \otimes 1 + 1 \otimes d_Z$ where~$X$ is as before and $Z: U \rightarrow W$ for some potential~$U$. The unit for the fusion product is the \textsl{invisible defect} $I_W:W(x')\rightarrow W(x)$, where we agree to always use primed notation for the copy of variables pertaining to the source of an endodefect. As a module it is given by $I_W = \bigwedge (\bigoplus_{i=1}^n \C[x,x'] \cdot \theta_i)$, i.\,e.~by the exterior algebra generated by~$n$ anticommuting variables~$\theta_i$ (sometimes interpreted as boundary fermions). For the twisted differential on $I_W$ we have $d_{I_W} = \sum_{i=1}^n [ (x_i-x'_i) \cdot \theta^*_i + \partial_{[i]}^{x,x'} W \cdot \theta_i]$ where we use the divided difference operators defined by 
\be\label{eq:DDO}
\partial_{[i]}^{x,x'} W = \frac{W(x'_1,\ldots,x'_{i-1}, x_i, \ldots, x_n) - W(x'_1,\ldots,x'_{i}, x_{i+1}, \ldots, x_n)}{x_i - x'_i}
\, . 
\ee
It is straightforward to verify that $\End(I_W)$ is isomorphic to $\C[x]/(\partial_i W)$, in line with the intuition that fields living on $I_W$ are nothing but bulk fields. 

The invisible defect is the unit for fusion in the sense that there are natural isomorphisms (up to homotopy, i.\,e.~up to BRST exact terms) $\lambda_X: I_V \otimes X \rightarrow X$ and $\rho_X: X \otimes I_W \rightarrow X$ that implement its $\lambda$eft and $\rho$ight action on~$X$. For example, $\lambda_X$ is the projection $I_V\rightarrow \C[z,z']$ to $\theta$-degree zero, followed by identifying $z'=z$, and analogously for~$\rho_X$. We may depict these maps and their inverses as follows: 
$$
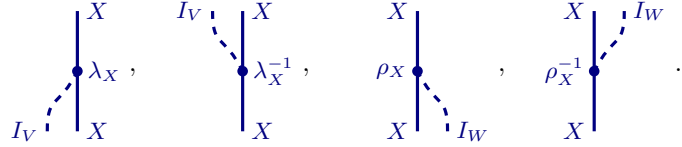
\begin{tikzpicture}[very thick,scale=1.0,color=blue!50!black, baseline=0cm]

\fill (0,0) circle (2.0pt) node[right] {{\small $\lambda_X$}};

\draw[dashed] (-0.4,-0.8) .. controls +(0,0.5) and +(-0.25,-0.5) .. (0,0);

\draw[line width=0] 
(0,-0.8) node[line width=0pt, right] (Xbottom) {{\small $X$}}
(0,0.8) node[line width=0pt, right] (Xtop) {{\small $X$}}
(-0.4,-0.8) node[line width=0pt, left] (I) {{\small $I_V$}};
\draw (0,-0.8) -- (0,0.8); 

\end{tikzpicture}
, \quad
\begin{tikzpicture}[very thick,scale=1.0,color=blue!50!black, baseline=0cm]

\fill (0,0) circle (2.0pt) node[right] {{\small $\lambda_X^{-1}$}};

\draw[dashed] (-0.4,0.8) .. controls +(0,-0.5) and +(-0.25,0.5) .. (0,0);

\draw[line width=0] 
(0,-0.8) node[line width=0pt, right] (Xbottom) {{\small $X$}}
(0,0.8) node[line width=0pt, right] (Xtop) {{\small $X$}}
(-0.4,0.8) node[line width=0pt, left] (I) {{\small $I_V$}};
\draw (0,-0.8) -- (0,0.8); 

\end{tikzpicture}
, \qquad 
\begin{tikzpicture}[very thick,scale=1.0,color=blue!50!black, baseline=0cm]

\fill (0,0) circle (2.0pt) node[left] {{\small $\rho_X$}};

\draw[dashed] (0.4,-0.8) .. controls +(0,0.5) and +(0.25,-0.5) .. (0,0);

\draw[line width=0] 
(0,-0.8) node[line width=0pt, left] (Xbottom) {{\small $X$}}
(0,0.8) node[line width=0pt, left] (Xtop) {{\small $X$}}
(0.4,-0.8) node[line width=0pt, right] (I) {{\small $I_W$}};
\draw (0,-0.8) -- (0,0.8); 

\end{tikzpicture}
, \quad 
\begin{tikzpicture}[very thick,scale=1.0,color=blue!50!black, baseline=0cm]

\fill (0,0) circle (2.0pt) node[left] {{\small $\rho_X^{-1}$}};

\draw[dashed] (0.4,0.8) .. controls +(0,-0.5) and +(0.25,0.5) .. (0,0);

\draw[line width=0] 
(0,-0.8) node[line width=0pt, left] (Xbottom) {{\small $X$}}
(0,0.8) node[line width=0pt, left] (Xtop) {{\small $X$}}
(0.4,0.8) node[line width=0pt, right] (I) {{\small $I_W$}};
\draw (0,-0.8) -- (0,0.8); 

\end{tikzpicture}
.
$$
More generally, any collection of fields sitting on defects adorning some worldsheet is naturally described by such a diagram. To determine a given diagram's `value', i.\,e.~the effective field inserted at the junction of all in- and outgoing defects, we read it from bottom to top (operator product) and from right to left (fusion product). 

Next we consider the defect~$X^\dagger$ that has the opposite orientation but otherwise imposes the same constraints on bulk fields as~$X$. For Landau-Ginzburg models we have $X^\dagger = X^\vee[n] \equiv \Hom_{\C[z,x]}(X, \C[z,x])[n]$ with twisted differential
$$
d_{X^\dagger} = \begin{pmatrix} 0 & (d_X^0)^\vee \\ - (d_X^1)^\vee & 0 \end{pmatrix}\![n]
\, , \quad
\text{if} \quad
d_X = \begin{pmatrix} 0 & d_X^1 \\ d_X^0 & 0\end{pmatrix}
. 
$$
For $\Psi \in \Hom(X,Y)$ the orientation reversed field is $\Psi^\dagger = \Psi^\vee[n] \in \Hom(Y^\dagger, X^\dagger)$. 

Oriented defects may `take a U-turn', so we expect canonical defect fields
\be\label{eq:tilde-adj}
\begin{tikzpicture}[thick,scale=0.5,color=blue!50!black, baseline=.18cm]
\draw[line width=0pt] 
(3,0) node[line width=0pt] (D) {}
(2,0) node[line width=0pt] (s) {}; 
\draw[redirected] (D) .. controls +(0,1) and +(0,1) .. (s);
\end{tikzpicture}
\equiv \widetilde\eval_X  
: X \otimes X^\dagger \longrightarrow I_V \, , \quad 
\begin{tikzpicture}[thick,scale=0.5,color=blue!50!black, baseline=-.35cm,rotate=180]
\draw[line width=0pt] 
(3,0) node[line width=0pt] (D) {}
(2,0) node[line width=0pt] (s) {}; 
\draw[directed] (D) .. controls +(0,1) and +(0,1) .. (s);
\end{tikzpicture}
\equiv \widetilde\coev_X : I_W \longrightarrow X^\dagger \otimes X 
\ee
where up- and downwards oriented lines are implictly labelled by~$X$ and $X^\dagger$, respectively, and we no longer display dashed lines for the invisible defect. Furthermore, by the topological nature of~$X$ we expect the identities
\be\label{eq:Zorro}
\begin{tikzpicture}[very thick,scale=0.7,color=blue!50!black, baseline=0cm]
\draw[line width=0] 
(1,1.25) node[line width=0pt] (A) {}
(-1,-1.25) node[line width=0pt] (A2) {}; 
\draw[redirected] (0,0) .. controls +(0,1) and +(0,1) .. (-1,0);
\draw[redirected] (1,0) .. controls +(0,-1) and +(0,-1) .. (0,0);
\draw (-1,0) -- (A2); 
\draw (1,0) -- (A); 
\end{tikzpicture}
=
\begin{tikzpicture}[very thick,scale=0.7,color=blue!50!black, baseline=0cm]
\draw[line width=0] 
(0,1.25) node[line width=0pt] (A) {}
(0,-1.25) node[line width=0pt] (A2) {}; 
\draw[
	decoration={markings, mark=at position 0.55 with {\arrow{>}}}, postaction={decorate}
	]
 (A2) -- (A); 
\end{tikzpicture}
\, , \qquad
\begin{tikzpicture}[very thick,scale=0.7,color=blue!50!black, baseline=0cm]
\draw[line width=0] 
(-1,1.25) node[line width=0pt] (A) {}
(1,-1.25) node[line width=0pt] (A2) {}; 
\draw[redirected] (0,0) .. controls +(0,-1) and +(0,-1) .. (-1,0);
\draw[redirected] (1,0) .. controls +(0,1) and +(0,1) .. (0,0);
\draw (-1,0) -- (A); 
\draw (1,0) -- (A2); 
\end{tikzpicture}
=
\begin{tikzpicture}[very thick,scale=0.7,color=blue!50!black, baseline=0cm]
\draw[line width=0] 
(0,1.25) node[line width=0pt] (A) {}
(0,-1.25) node[line width=0pt] (A2) {}; 
\draw[
	decoration={markings, mark=at position 0.5 with {\arrow{>}}}, postaction={decorate}
	]
 (A) -- (A2); 
\end{tikzpicture} 
\ee
to hold. While the diagrams on either side of each equality are certainly isotopic, \eqref{eq:Zorro} does impose nontrivial conditions on the maps~\eqref{eq:tilde-adj}; e.\,g.~the second identity reads $\rho_{X^\dagger} \circ (1_{X^\dagger} \otimes \widetilde\eval_X)Ê\circ (\widetilde\coev_X \otimes 1_{X^\dagger}) \circ \lambda_{X^\dagger}^{-1} = 1_{X^\dagger}$ in clumsy diagram-free notation. We say that two defect fields $\widetilde\eval_X, \widetilde\coev_X$ exhibit~$X^\dagger$ as the \textsl{right adjoint} of~$X$ if the \textsl{Zorro moves}~\eqref{eq:Zorro} hold. This generalises the adjunction for a finite-dimensional vector space~$V$ and its dual $V^{\dagger}$ encoded in the standard evaluation and coevaluation maps defined by $\widetilde\eval_V(v\otimes \alpha) = \alpha(v)$ and $\widetilde\coev_V: 1\mapsto \sum_i e_i^* \otimes e_i$ for any basis $\{e_i\}$ of~$V$. Analogously, ${}^\dagger X = X^\vee[m]$ is \textsl{left adjoint} to~$X$ if there are defect fields 
\be\label{eq:adj}
\begin{tikzpicture}[thick,scale=0.5,color=blue!50!black, baseline=.18cm]
\draw[line width=0pt] 
(3,0) node[line width=0pt] (D) {}
(2,0) node[line width=0pt] (s) {}; 
\draw[directed] (D) .. controls +(0,1) and +(0,1) .. (s);
\end{tikzpicture}
 \equiv \eval_X: {}^\dagger X \otimes X \rightarrow I_W 
 \, , \quad
\begin{tikzpicture}[thick,scale=0.5,color=blue!50!black, baseline=-.35cm,rotate=180]
\draw[line width=0pt] 
(3,0) node[line width=0pt] (D) {}
(2,0) node[line width=0pt] (s) {}; 
\draw[redirected] (D) .. controls +(0,1) and +(0,1) .. (s);
\end{tikzpicture}
 \equiv \coev_X: I_V \rightarrow X \otimes {}^\dagger X
 \ee
 satisfying their versions of Zorro moves: 
 \be\label{eq:Zorro2}
\begin{tikzpicture}[very thick,scale=0.7,color=blue!50!black, baseline=0cm]
\draw[line width=0] 
(-1,1.25) node[line width=0pt] (A) {}
(1,-1.25) node[line width=0pt] (A2) {}; 
\draw[directed] (0,0) .. controls +(0,-1) and +(0,-1) .. (-1,0);
\draw[directed] (1,0) .. controls +(0,1) and +(0,1) .. (0,0);
\draw (-1,0) -- (A); 
\draw (1,0) -- (A2); 
\end{tikzpicture}
=
\begin{tikzpicture}[very thick,scale=0.7,color=blue!50!black, baseline=0cm]
\draw[line width=0] 
(0,1.25) node[line width=0pt] (A) {}
(0,-1.25) node[line width=0pt] (A2) {}; 
\draw[
	decoration={markings, mark=at position 0.55 with {\arrow{>}}}, postaction={decorate}
	]
(A2) -- (A); 
\end{tikzpicture}
\, , \quad
\begin{tikzpicture}[very thick,scale=0.7,color=blue!50!black, baseline=0cm]
\draw[line width=0] 
(1,1.25) node[line width=0pt] (A) {}
(-1,-1.25) node[line width=0pt] (A2) {}; 
\draw[directed] (0,0) .. controls +(0,1) and +(0,1) .. (-1,0);
\draw[directed] (1,0) .. controls +(0,-1) and +(0,-1) .. (0,0);
\draw (-1,0) -- (A2); 
\draw (1,0) -- (A); 
\end{tikzpicture}
=
\begin{tikzpicture}[very thick,scale=0.7,color=blue!50!black, baseline=0cm]
\draw[line width=0] 
(0,1.25) node[line width=0pt] (A) {}
(0,-1.25) node[line width=0pt] (A2) {}; 
\draw[
	decoration={markings, mark=at position 0.5 with {\arrow{>}}}, postaction={decorate}
	]
(A) -- (A2); 
\end{tikzpicture} \, . 
\ee

The main theorem, quoted below in its explicit form, gives expressions for the adjunction maps~\eqref{eq:tilde-adj}, \eqref{eq:adj} as well as the inverses $\lambda_X^{-1}, \rho_X^{-1}$. For practical purposes it is important to note that the formulas are all expressed concretely in terms of matrix multiplication, taking derivatives, and `integrating' -- where we recall that residues are computed by the rules
\[
\Res \left[ \frac{\operatorname{d}\! x }{x_1^{a_1} ,\ldots, x_n^{a_n}} \right] = \delta_{a_1,1}\ldots \delta_{a_n,1} 
\quad \text{and} \quad 
\Res \left[ \frac{h \, \operatorname{d}\! x }{f_1, \ldots, f_n} \right] 
= 
\Res \left[ \frac{\det(C) h \, \operatorname{d}\! x }{g_1, \ldots, g_n} \right]
\]
if $g_i = \sum_{j=1}^n C_{ij} f_j$ for polynomials $h, f_i, C_{ij}$. The upshot is that the correlator of any planar defect-decorated worldsheet can be straightforwardly (and often gainfully) computed by viewing it as a diagram made up of the maps~\eqref{eq:tilde-adj}, \eqref{eq:adj}, $\lambda^{\pm 1}, \rho^{\pm 1}$ and whatever other defect fields may be present. We will return to this point and discuss several examples in the next section. 

\begin{theorem}[\cite{cm1208.1481}]
Any matrix factorisation~$X$ of $V(z_1,\ldots,z_m)-W(x_1,\ldots,x_n)$ has left and right adjoints. The associated structure maps have the following explicit presentations: 
\begin{align*}
\widetilde\eval_X( e_j \otimes e_i^* ) 
& = \sum_{l \geqslant 0} \sum_{a_1 < \cdots < a_l} (-1)^{l + (n+1)|e_j|} \, \theta_{a_1} \ldots \theta_{a_l} \\
& \qquad \cdot \Res \left[ \frac{ \big\{ \partial^{z,z'}_{[a_l]} d_X  \ldots \partial^{z,z'}_{[a_1]} d_X \, \Lambda^{(x)} \big\}_{ij} \, \operatorname{d}\!x }{\partial_{x_1}W, \ldots, \partial_{x_n} W} \right] , \\
\eval_X( e_i^* \otimes e_j )
&  = \sum_{l \geqslant 0} \sum_{a_1 < \cdots < a_l} (-1)^{\binom{l}{2}+l|e_j|} \, \theta_{a_1} \ldots \theta_{a_l} \\
& \qquad \cdot \Res \left[ \frac{ \big\{ \Lambda^{(z)} \, \partial^{x,x'}_{[a_1]} d_X  \ldots \partial^{x,x'}_{[a_l]} d_X \big\}_{ij}  \, \operatorname{d}\!z }{\partial_{z_1}V, \ldots, \partial_{z_m} V} \right] , \\
\widetilde\coev_X( \bar\gamma ) 
& = \sum_{i,j} (-1)^{(\bar r+1)|e_j| + s_n} \left\{ \partial^{x,x'}_{[\bar b_{\bar r}]}(d_X) \ldots \partial^{x,x'}_{[\bar b_1]}(d_X) \right\}_{ji}  e_i^* \otimes e_j \, , \\
\coev_X(\gamma) 
& = \sum_{i,j} (-1)^{\binom{r+1}{2} + mr + s_m} \left\{ \partial^{z,z'}_{[b_1]}d_X \ldots \partial^{z,z'}_{[b_r]}d_X \right\}_{ij} e_{i} \otimes e_j^* \, , \\
\lambda^{-1}_X(e_i) 
& = \sum_{l \geqslant 0} \sum_{a_1 < \cdots < a_l} \sum_j
\theta_{a_1} \ldots \theta_{a_l} 
\left\{ \partial^{z,z'}_{[a_l]}d_X \ldots \partial^{z,z'}_{[a_1]}d_X \right\}_{ji} \otimes e_{j}
\, , \\
\rho^{-1}_X(e_i) 
& = \sum_{l \geqslant 0} \sum_{a_1 < \cdots < a_l} \sum_j (-1)^{\binom{l}{2} + l|e_i|} e_j \otimes \left\{ \partial^{x,x'}_{[a_1]}d_X \ldots \partial^{x,x'}_{[a_l]}d_X \right\}_{ji} \theta_{a_1} \ldots \theta_{a_l}
\end{align*}
where $\{e_i\}$ is a basis of~$X$, $\Lambda^{(x)} := (-1)^n \partial_{x_1}d_X\ldots \partial_{x_n}d_X$, $\Lambda^{(z)} := \partial_{z_1}d_X\ldots \partial_{z_m}d_X$, 
and $b_i, \bar b_{\bar\jmath}$ and $s_n, s_m\in\Z_2$ are uniquely determined by requiring that $b_1 < \cdots < b_r$, $\bar b_1 < \cdots < \bar b_{\bar r}$, and $\bar{\gamma} \theta_{\bar b_1} \ldots \theta_{\bar b_{\bar r}} = (-1)^{s_n} \theta_1 \ldots \theta_n$ and $\gamma \theta_{b_1} \ldots \theta_{b_r} = (-1)^{s_m} \theta_1 \ldots \theta_m$.
\end{theorem}

Let us briefly comment on the proof. Its first ingredient is homological perturbation~\cite{c0403266}, which allows us to construct $\lambda_X^{-1}, \rho_X^{-1}$ by viewing parts of the twisted differential of, say, $I_V \otimes X$ as a perturbation~\cite[Sect.\,4]{cm1208.1481}. Then using noncommutative forms for another natural presentation of the invisible defect leads to a conceptually solid description of $\lambda_X^{-1}, \rho_X^{-1}$ in terms of `associative Atiyah classes'~\cite[Sect.\,3]{cm1208.1481}. Computing them explicitly gives rise to the divided difference operators $\partial_{[i]}^{x,x'}d_X$ in the formulas above. 

The coevaluation is the image of the identity under the map~\cite[Eq.\,(5.11)]{cm1208.1481}
$$
\xymatrix{%
\Hom(X,X)
\cong X^\vee \otimes X 
\ar[rr]^-{{\rho'}_{X^\vee \otimes X}^{-1}} && 
 X^\vee \otimes X \otimes I'_W
\cong \Hom(I_W, X^\vee[n] \otimes X)
}%
$$
where $I'_W$ denotes $I_W$ but with twisted differential $\sum_{i=1}^n [ (x_i-x'_i) \cdot \theta^*_i - \partial_{[i]}^{x,x'} W \cdot \theta_i]$. Note that the nontrivial part ${\rho'}_{X^\vee \otimes X}^{-1}$ is an isomorphism only up to homotopy. Similarly, the evaluation is constructed by lifting the Kapustin-Li pairing~\cite{kl0305, hl0404} $X\otimes_{\C[x]} X^\dagger \rightarrow \C[z]$ using homological perturbation~\cite[Sect.\,5.2]{cm1208.1481}. 

Proving that the above adjunction maps really satisfy the Zorro moves (up to homotopy) is one of the central results of~\cite[Sect.\,6]{cm1208.1481}. To give a taste we work out a simple example that boils down to properties of Pauli matrices in Appendix~\ref{app:ZorroDirac}. The general case essentially amounts to artfully manipulating Atiyah classes inside the residue and supertrace. 

\medskip

Finally, we collect a few useful identities. Together with the Zorro moves and the expressions in our theorem they are all we need to compute any correlator of defect fields. 

\begin{proposition}[\cite{cm1208.1481}]
For $\Psi \in \Hom(X,Y)$ and composable $X,Z$ we have 
\begin{align*}
& 
\begin{tikzpicture}[very thick,scale=0.7,color=blue!50!black, baseline=0cm]
\draw[line width=0] 
(0,1.25) node[line width=0pt] (A) {}
(0,-1.25) node[line width=0pt] (A2) {}; 
\draw (A2) -- (A); 
\fill (0,0) circle (2.5pt) node[left] {{\small $\Psi^\dagger$}};
\draw[line width=0] 
(0,1.25) node[line width=0pt, left] (X) {{\small $X^\dagger$}}
(0,-1.25) node[line width=0pt, left] (Y) {{\small $Y^\dagger$}};
\end{tikzpicture}
=
\begin{tikzpicture}[very thick,scale=0.7,color=blue!50!black, baseline=0cm]
\draw[line width=0] 
(-1,1.25) node[line width=0pt] (A) {}
(1,-1.25) node[line width=0pt] (A2) {}; 
\draw[redirected] (0,0) .. controls +(0,-1) and +(0,-1) .. (-1,0);
\draw[redirected] (1,0) .. controls +(0,1) and +(0,1) .. (0,0);
\draw (-1,0) -- (A); 
\draw (1,0) -- (A2); 
\fill (0,0) circle (2.5pt) node[left] {{\small $\Psi$}};
\draw[line width=0] 
(-1,1.25) node[line width=0pt, left] (X) {{\small $X^\dagger$}}
(2,-1.25) node[line width=0pt, left] (Y) {{\small $Y^\dagger$}};
\end{tikzpicture}
\, , \quad
\begin{tikzpicture}[very thick,scale=0.7,color=blue!50!black, baseline=.4cm]
\draw[line width=0pt] 
(3,0.5) node[line width=0pt] (D) {}
(2,0.5) node[line width=0pt] (s) {}; 
\draw[redirected] (D) .. controls +(0,-1) and +(0,-1) .. (s);
\fill (2,0.5) circle (2.5pt) node[right] {{\small $\Psi^\dagger$}};
\draw (3,0.35) -- (3,1)
node[above] {{{\small${}^\dagger X$}}};
\draw (2,0.35) -- (2,1)
node[above] {{{\small$Y$}}};
\end{tikzpicture} 
=
\begin{tikzpicture}[very thick,scale=0.7,color=blue!50!black, baseline=.4cm]
\draw[line width=0pt] 
(3,0.5) node[line width=0pt] (D) {}
(2,0.5) node[line width=0pt] (s) {}; 
\draw[redirected] (D) .. controls +(0,-1) and +(0,-1) .. (s);
\fill (3,0.5) circle (2.5pt) node[left] {{\small $\Psi$}};
\draw (3,0.35) -- (3,1)
node[above] {{{\small${}^\dagger X$}}};
\draw (2,0.35) -- (2,1)
node[above] {{{\small$Y$}}};
\end{tikzpicture} 
\, , \quad
\begin{tikzpicture}[very thick,scale=0.7,color=blue!50!black, baseline=.4cm]
\draw[line width=0pt] 
(3,0.5) node[line width=0pt] (D) {}
(2,0.5) node[line width=0pt] (s) {}; 
\draw[redirected] (D) .. controls +(0,1) and +(0,1) .. (s);
\fill (3,0.5) circle (2.5pt) node[left] {{\small $\Psi^\dagger$}};
\draw (3,0.65) -- (3,0)
node[below] {{{\small$X\vphantom{{}^\dagger Y}$}}};
\draw (2,0.65) -- (2,0)
node[below] {{{\small${}^\dagger Y$}}};
\end{tikzpicture} 
=
\begin{tikzpicture}[very thick,scale=0.7,color=blue!50!black, baseline=.4cm]
\draw[line width=0pt] 
(3,0.5) node[line width=0pt] (D) {}
(2,0.5) node[line width=0pt] (s) {}; 
\draw[redirected] (D) .. controls +(0,1) and +(0,1) .. (s);
\fill (2,0.5) circle (2.5pt) node[right] {{\small $\Psi$}};
\draw (3,0.65) -- (3,0)
node[below] {{{\small$X\vphantom{{}^\dagger Y}$}}};
\draw (2,0.65) -- (2,0)
node[below] {{{\small${}^\dagger Y$}}};
\end{tikzpicture} 
\, ,
\\
& 
\begin{tikzpicture}[very thick,scale=0.6,color=blue!50!black, baseline, rotate=180]
\draw[line width=1pt] 
(2,-2) node[line width=0pt, right] (Y) {{\small $Z^\dagger$}}
(3,-2) node[line width=0pt, right] (X) {{\small $X^\dagger$}}
(-1,3) node[line width=0pt, left] (XY) {{\small $(Z\otimes X)^\dagger$}}; 
\draw[redirected] (1,0) .. controls +(0,1) and +(0,1) .. (2,0);
\draw[redirected] (0,0) .. controls +(0,2) and +(0,2) .. (3,0);
\draw[redirected] (-1,0) .. controls +(0,-1) and +(0,-1) .. (0.5,0);
\draw (2,0) -- (2,-2);
\draw (3,0) -- (3,-2);
\draw[dotted] (0,0) -- (1,0);
\draw (-1,0) -- (-1,3);
\end{tikzpicture}
\, \simeq \, 
\begin{tikzpicture}[very thick,scale=0.6,color=blue!50!black, baseline, rotate=180]
\draw[line width=1pt] 
(-1,-2) node[line width=0pt, right] (X) {{\small $Z^\dagger$}}
(0,-2) node[line width=0pt, right] (Y) {{\small $X^\dagger$}}
(2,3) node[line width=0pt, right] (XY) {{\small $(Z\otimes X)^\dagger$}}; 
\draw[redirected] (0,0) .. controls +(0,1) and +(0,1) .. (-1,0);
\draw[redirected] (1,0) .. controls +(0,2) and +(0,2) .. (-2,0);
\draw[redirected] (2,0) .. controls +(0,-1) and +(0,-1) .. (0.5,0);
\draw (-1,0) -- (-1,-2);
\draw (-2,0) -- (-2,-2);
\draw[dotted] (0,0) -- (1,0);
\draw (2,0) -- (2,3);
\end{tikzpicture}
\, , \quad
\begin{tikzpicture}[very thick,scale=0.6,color=blue!50!black, baseline, rotate=180]
\draw[line width=1pt] 
(-2,-2) node[line width=0pt, left] (X) {{\small ${}^\dagger Z$}}
(-1,-2) node[line width=0pt, left] (Y) {{\small ${}^\dagger X$}}
(2,3) node[line width=0pt, right] (XY) {{\small ${}^\dagger (Z\otimes X)$}}; 
\draw[redirected] (0,0) .. controls +(0,1) and +(0,1) .. (-1,0);
\draw[redirected] (1,0) .. controls +(0,2) and +(0,2) .. (-2,0);
\draw[redirected] (2,0) .. controls +(0,-1) and +(0,-1) .. (0.5,0);
\draw (-1,0) -- (-1,-2);
\draw (-2,0) -- (-2,-2);
\draw[dotted] (0,0) -- (1,0);
\draw (2,0) -- (2,3);
\end{tikzpicture}
\, \simeq  \,
\begin{tikzpicture}[very thick,scale=0.6,color=blue!50!black, baseline, rotate=180]
\draw[line width=1pt] 
(1,-2) node[line width=0pt, left] (Y) {{\small ${}^\dagger Z$}}
(2,-2) node[line width=0pt, left] (X) {{\small ${}^\dagger X$}}
(-1,3) node[line width=0pt, left] (XY) {{\small ${}^\dagger (Z\otimes X)$}}; 
\draw[redirected] (1,0) .. controls +(0,1) and +(0,1) .. (2,0);
\draw[redirected] (0,0) .. controls +(0,2) and +(0,2) .. (3,0);
\draw[redirected] (-1,0) .. controls +(0,-1) and +(0,-1) .. (0.5,0);
\draw (2,0) -- (2,-2);
\draw (3,0) -- (3,-2);
\draw[dotted] (0,0) -- (1,0);
\draw (-1,0) -- (-1,3);
\end{tikzpicture}
\, ,
\end{align*}
hiding certain signs (explained at length in~\cite[Sect.\,7]{cm1208.1481}) in the symbol~$\simeq$. 
\end{proposition}

\section{Applications}\label{sec:apps}

We now illustrate how the general results collected in the previous section are put to use. As already stated, the explicit expressions in our theorem allow us to compute arbitrary correlators of planar worldsheets with defects, for which we give two examples. Another application is to the theory of generalised orbifolds. 

\subsection*{Defect actions}

Let us fix a defect $X:W\rightarrow V$, a defect field $\Psi\in \End(X)$ and a bulk field $\phi\in\End(I_W)$. We will explain that for the associated \textsl{defect action on bulk fields} we have
\be\label{eq:defectaction}
\begin{tikzpicture}[ thick,scale=0.37,color=blue!50!black, baseline]
\nicenotpalecolourscheme (0,0) circle (3.5);
\fill (2.2,-2.2) circle (0pt) node[white] {{\small$V$}};
\nicedashedcolourscheme (0,0) circle (2);
\fill (1.1,-1.1) circle (0pt) node[white] {{\small$W$}};
\fill (180:2) circle (3.9pt) node[left] {{\small$\Psi$}};
\fill (45:1.95) circle (0pt) node[right] {{\small$X$}};
\draw (0,0) circle (2);
\draw[<-,  thick] (0.100,2) -- (-0.101,2) node[above] {}; 
\draw[<-,  thick] (-0.100,-2) -- (0.101,-2) node[below] {}; 
\fill (135:0) circle (3.9pt) node[right] {{\small$\phi$}};
\end{tikzpicture} 
 =
\begin{tikzpicture}[ thick,scale=0.37,color=blue!50!black, baseline]
\nicenotpalecolourscheme (0,0) circle (3.5);
\fill (2.2,-2.2) circle (0pt) node[white] {{\small$V$}};
\nicedashedcolourscheme (0,0) circle (2);
\fill (1.1,-1.1) circle (0pt) node[white] {{\small$W$}};
\fill (180:2) circle (3.9pt) node[left] {{\small$\Psi$}};
\fill (45:1.95) circle (0pt) node[right] {{\small$X$}};
\draw (0,0) circle (2);
\draw[<-,  thick] (0.100,2) -- (-0.101,2) node[above] {}; 
\draw[<-,  thick] (-0.100,-2) -- (0.101,-2) node[below] {}; 
\fill (135:0) circle (3.9pt) node[right] {{\small$\phi$}};
\fill (130:2) circle (3.9pt) node[left] {{\scriptsize$\rho_{X}$}};
\fill (230:2) circle (3.9pt) node[left] {{\scriptsize$\rho^{-1}_{X}$}};
\fill (84:2.45) circle (0pt) node[left] {{\scriptsize$\widetilde\eval_X$}};
\fill (-82:2.45) circle (0pt) node[left] {{\scriptsize$\coev_X$}};
\draw[dashed] (135:0) .. controls +(0,1) and +(0.25,-1) .. (130:2);
\draw[dashed] (135:0) .. controls +(0,-1) and +(0.25,1) .. (230:2);
\draw[dashed] (270:2) -- (270:3.3);
\draw[dashed] (90:2) -- (90:3.3);
\end{tikzpicture} 
= 
(-1)^{{m+1}\choose 2}
\Res\!\! \left[ \frac{\phi \str \!\left( \Psi \big( \prod_i \partial_{x_i} d_X\big) \! \big( \prod_j \partial_{z_j} d_X \big)\! \right) \! \operatorname{d}\! x }{\partial_{x_1} W, \ldots, \partial_{x_n} W} \right]
. 
\ee
Here the left-hand side shows the physical picture of the defect~$X$, decorated by~$\Psi$, wrapping around the bulk field~$\phi$ of the theory~$W$. Because of the topological nature of the situation, $X$ may wrap~$\phi$ as tightly as we please; the limit of the defect line collapsing on the bulk field (of the inner theory~$W$) is effectively described by a new bulk field (of the outer theory~$V$) that we call $\mathcal D_X^\Psi(\phi)$. It is precisely this effective bulk field that is computed in~\eqref{eq:defectaction}. 

The first step in determining the defect action in~\eqref{eq:defectaction} is to translate the leftmost physical picture into rigorous mathematical language. Given our diagrammatics this is straightforward: all we have to do is view bulk fields as endomorphisms of the invisible defect ($\phi \in \End(I_W)$, $\mathcal D_X^\Psi(\phi) \in \End(I_V)$), make the latter and its action on other defects visible (i.\,e.~insert $\rho_X, \rho_X^{-1}$ in our example), and label cups and caps of defect lines by appropriate adjunction maps ($\widetilde\eval_X$, $\coev_X$). This is done in the first step of~\eqref{eq:defectaction}; reading the resulting diagram from bottom to top and from right to left produces the new bulk field $\mathcal D_X^\Psi(\phi)$ in theory~$V$. To compute it explicitly, in the second step of~\eqref{eq:defectaction} we simply plug in our expressions for $\rho^{\pm 1}, \widetilde\eval, \coev$ from the previous section. After a short calculation~\cite[Sect.\,8]{cm1208.1481} this leads to the residue formula on the right-hand side. Note that it gives a precise meaning to `integrating out' the $x$-dependent degrees of freedom in theory~$W$. It is clear that any planar configuration of defects with field insertions can be computed in this way. 

The defect action~\eqref{eq:defectaction} has several special cases: 
\begin{itemize}
\item 
If we set both~$\phi$ and~$\Psi$ equal to the respective identities, i.\,e.~if we simply consider an empty defect bubble labelled by~$X$, then by definition we obtain the \textsl{(right) quantum dimension} of~$X$. 
\item 
If $V=0$, i.\,e.~if the outer theory is trivial and the defect becomes a boundary condition, then~\eqref{eq:defectaction} recovers the Kapustin-Li \textsl{disc correlator} of~\cite{kl0305, hl0404}. 
\item 
If $W=0$, i.\,e.~if the inner theory is trivial, \eqref{eq:defectaction} describes the \textsl{boundary-bulk map} $\Psi \mapsto (-1)^{{m+1}\choose 2} \str(\Psi\, \partial_{z_1}d_X \ldots \partial_{z_m}d_X)$ of~\cite{kr0405232}. If we further restrict to $\Psi = 1_X$, we obtain what is traditionally called the \textsl{boundary state} or \textsl{Chern character} of~$X$. 
\end{itemize}

\subsection*{Cardy condition}

We just saw that disc correlators and boundary-bulk maps can be neatly formulated in defect language. Actually the complete structure of open/closed topological field theory (TFT) naturally fits into this framework~\cite[Sect.\,9]{cm1208.1481}. As a second example of our theorem's practical uses we now recall how it is used to give a `one-line proof' of the Cardy condition. 

The Cardy condition is most familiar in two-dimensional conformal field theory (CFT), where it is a consequence of open/closed duality and can be derived by evaluating a cylinder amplitude in two different ways, see e.\,g.~\cite{blumenhagenbook}. It asserts that the overlap of two boundary states equals the associated open sector partition function which is computed as a certain trace. Similarly, the TFT version of the Cardy condition says that the two-point bulk correlator of two images under the boundary-bulk map is the same as a certain supertrace in the boundary sector. It can be considered the most `quantum' gluing axiom as it originates from inspection of a genus-one worldsheet. Another incarnation of its importance is that in the case of B-twisted sigma models the Cardy condition manifests itself as a generalisation of the Hirzebruch-Riemann-Roch theorem~\cite{ct1007.2679}. 

We will now first formulate the Cardy condition for Landau-Ginzburg models (originally proven in~\cite{pv1002.2116, dm1102.2957,buchweitzstratten}) and then explain how to effortlessly derive it from the defect perspective advocated in this note. 

Let $X,Y$ be two matrix factorisations of $W(x_1,\ldots,x_n)$ and $\Phi \in \End(X)$, $\Psi \in \End(Y)$. Then the Cardy condition states
\be\label{eq:Cardy}
(-1)^{{n+1}\choose 2} 
\Res \left[ \frac{\str\left( \Phi \, \partial_{x_1} d_X \ldots \partial_{x_n} d_X\right) \str \left( \Psi \, \partial_{x_1} d_Y \ldots \partial_{x_n} d_Y \right)\operatorname{d}\! x }{\partial_{x_1} W, \ldots, \partial_{x_n} W} \right] 
= \str\left( {}_\Psi m_\Phi \right) 
\ee
where ${}_\Psi m_\Phi$ is the operator on the open string space $\Hom(X,Y)$ that precomposes with~$\Phi$ and postcomposes with~$\Psi$ (so in particular ${}_{1_Y} m_{1_X} = 1_{\Hom(X,Y)}$ and the supertrace becomes a simple index). 

In order to argue for~\eqref{eq:Cardy} we think of a cylinder with boundary conditions as an annulus correlator. Accordingly, we claim that the proof of the Cardy condition is contained in the following identities: 
\be\label{eq:Cardy2}
(-1)^{{n+1}\choose 2} \,  
\begin{tikzpicture}[very thick,scale=0.60,color=blue!50!black, baseline,>=stealth]
\nicepalecolourscheme (0,0) circle (2);
\fill (1.5,0) circle (0pt) node[white] {{\small$W$}};
\draw (0,0) circle (2);
\draw[->, very thick] (-0.100,2) -- (-0.101,2) node[above] {}; 
\draw[->, very thick] (0.100,-2) -- (0.101,-2) node[below] {}; 
\fill (45:2) circle (2.5pt) node[right] {{\small$\Psi$}};
\fill (45:0) circle (2.5pt) node[below] {{\small$\str(\Phi\Lambda_X)$}};
\fill (0:2.8) circle (0pt) node[left] {{\small$Y$}};
\end{tikzpicture} 
=
\begin{tikzpicture}[very thick,scale=0.60,color=blue!50!black, baseline,>=stealth]
\nicepalecolourscheme (0,0) circle (2);
\nicereallynocolourscheme (0,0) circle (1);
\fill (1.5,0) circle (0pt) node[white] {{\small$W$}};
\draw (0,0) circle (2);
\draw[->, very thick] (-0.100,2) -- (-0.101,2) node[above] {}; 
\fill (45:2) circle (2.5pt) node[right] {{\small$\Psi$}};
\draw (0,0) circle (1);
\draw[->, very thick] (0.100,1) -- (0.101,1) node[above] {}; 
\fill (135:1) circle (2.5pt) node[left] {{\small$\Phi$}};
\fill (180:1.0) circle (0pt) node[right] {{\small$X$}};
\fill (0:2.8) circle (0pt) node[left] {{\small$Y$}};
\end{tikzpicture} 
=
\begin{tikzpicture}[very thick,scale=0.60,color=blue!50!black, baseline,>=stealth]
\draw[ultra thick] (0,0) circle (2);
\draw[->, ultra thick] (-0.100,2) -- (-0.101,2) node[above] {}; 
\draw[->, ultra thick] (0.100,-2) -- (0.101,-2) node[below] {}; 
\fill (45:2) circle (3.3pt) node[right] {{\small$\Phi^\dagger\otimes\Psi$}};
\fill (0:2.0) circle (0pt) node[right] {{\small$X^{\dagger} \otimes Y$}};
\end{tikzpicture} 
\ee
To make sense of this we start from the annulus diagram in the middle. Since we are dealing with a topological theory, the size of the inner $X$-boundary does not matter and may be shrunk to zero. This is the left equality, where we have used the knowledge from our first example, namely that the boundary-bulk map is given by $(-1)^{{n+1}\choose 2} \str(\Phi\Lambda_X)$ with $\Lambda_X = \partial_{x_1} d_X \ldots \partial_{x_n} d_X$. Next we use the other special case of~\eqref{eq:defectaction}, to wit the one with trivial outer theory, to immediately find that the disc correlator on the left-hand side of~\eqref{eq:Cardy2} is precisely given by the left-hand side of~\eqref{eq:Cardy}. To see that the right-hand side of \eqref{eq:Cardy} matches the right-hand side of \eqref{eq:Cardy2}, we note that the second equality in the latter comes about by expanding the inner boundary until it fuses with the outer boundary (where on the technical level use has to be made of a property called `pivotality', see~\cite[Sect.\,7]{cm1208.1481}). One can then convince oneself that the right-hand side of~\eqref{eq:Cardy2} is indeed given by $\str({}_\Psi m_\Phi)$, thus concluding our `one-line proof'  of the Cardy condition.

\subsection*{Generalised orbifolds}

As a further application of our explicit control over adjunctions in Landau-Ginzburg models we mention generalised orbifolds. Recall that given a finite symmetry group~$G$ of some two-dimensional field theory~$T$, correlators in the conventional orbifold theory $T^G$ can be computed as correlators in the original theory~$T$, but with a network of defects $A_G$ (implementing the group action) covering the worldsheet. It was first realised in~\cite{ffrs0909.5013} (within the framework of rational CFT) that this construction can be generalised by allowing \textsl{any} defect~$A$ that shares certain crucial properties with $A_G$ (to wit, $A$ must be a `separable symmetric Frobenius algebra', see e.\,g.~\cite[Sect.\,2.2]{genorb}). Among many other consequences, this means that any two rational CFTs with identical chiral algebras and central charges (that have a unique vacuum and nondegenerate two-point correlators) are generalised orbifolds of one another; this is in particular true for all minimal models with ADE-type modular invariants. 

These results were developed into a general theory of \textsl{orbifold completion} for arbitrary two-dimensional TFTs with defects in~\cite{genorb}. Unorbifolded theories~$T$ are `completed' by considering all pairs $(T,A)$ where $A:T\rightarrow T$ is a separable symmetric Frobenius algebra. Such pairs are called \textsl{generalised orbifolds}. The original theory~$T$ is identified with $(T,I_T)$, and ordinary orbifolds are the special cases $(T,A_G)\equiv T^G$. 

Much can be said already on this general level; to keep the discussion brief we shall only mention one central result of~\cite{genorb}. Let $X:T\rightarrow T'$ be a defect between theories $T,T'$ that has invertible quantum dimension. Then with $A_X:=X^\dagger \otimes X$ we have an equivalence of theories 
\be\label{eq:genorbequ}
T'\equiv (T',I_{T'})\cong (T,A_X) \, . 
\ee 
On the level of correlators, the idea behind this construction is illustrated by 
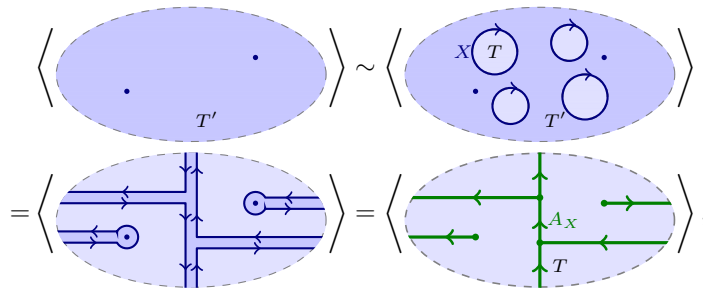
\begin{align*}
& 
\left\langle
\begin{tikzpicture}[thick,scale=0.3,color=blue!50!black, baseline]
\clip (0,0) ellipse (6cm and 3cm);
\nicedashedcolourscheme (0,0) ellipse (6cm and 3cm);
%
\draw (0.7,-2) [black] node {{\scriptsize $T'$}};
\fill (-2.85,-0.75) circle (3.3pt) node[below] {};
\fill (2.85,0.75) circle (3.3pt) node[above] {};
\end{tikzpicture}
\right\rangle
\sim
\left\langle
\begin{tikzpicture}[thick,scale=0.3,color=blue!50!black, baseline]
\clip (0,0) ellipse (6cm and 3cm);
\nicedashedcolourscheme (0,0) ellipse (6cm and 3cm);
%
\coordinate (circ1) at ($ (-2,1) $);
\nicedashedpalecolourscheme (circ1) circle (1);
\draw[
	decoration={markings, mark=at position 0.25 with {\arrow{<}}}, postaction={decorate}
	] 
	(circ1) circle (1);
\coordinate (circ1) at ($ (2,-1) $);
\nicedashedpalecolourscheme (circ1) circle (1);
\draw[
	decoration={markings, mark=at position 0.25 with {\arrow{<}}}, postaction={decorate}
	] 
	(circ1) circle (1);
\coordinate (circ2) at ($ (1.3,1.4) $);
\nicedashedpalecolourscheme (circ2) circle (0.8);
\draw[
	decoration={markings, mark=at position 0.25 with {\arrow{<}}}, postaction={decorate}
	] 
	(circ2) circle (0.8);
\coordinate (circ2) at ($ (-1.3,-1.4) $);
\nicedashedpalecolourscheme (circ2) circle (0.8);
\draw[
	decoration={markings, mark=at position 0.25 with {\arrow{<}}}, postaction={decorate}
	] 
	(circ2) circle (0.8);
%
\fill (-2.85,-0.75) circle (3.3pt) node[left] {};
\fill (2.85,0.75) circle (3.3pt) node[right] {};
%
\draw (0,0) [black] node {};
\draw (0,-2.25) [black] node {};
\draw (0.7,-2) [black] node {{\scriptsize $T'$}};
\draw (-2,1) [black] node {{\scriptsize $T$}};
\draw (-3.4,1) node {{\scriptsize $X$}};
\end{tikzpicture}
\right\rangle
\\
= & 
\left\langle
\begin{tikzpicture}[thick,scale=0.3,color=blue!50!black, baseline, rounded corners=0.1pt]
\clip (0,0) ellipse (6cm and 3cm);
\nicedashedpalecolourscheme (0,0) ellipse (6cm and 3cm);
\coordinate (circ1) at ($ (-150:0.5) + (-2.85,-0.75) $);
\coordinate (circ2) at ($ (-150:0.5) + (-5.85,-0.75) $);
\coordinate (circ3) at ($ (150:0.5) + (-2.85,-0.75) $);
\coordinate (circ4) at ($ (150:0.5) + (-5.85,-0.75) $);
\filldraw
	[
	fill=blue!22, 
	decoration={markings, mark=at position 0.23 with {\arrow{>}},
					mark=at position 0.745 with {\arrow{>}}
					}, postaction={decorate}
	] 
	(circ2) -- (circ1) arc (-150:150:0.5) -- (circ3) -- (circ4) -- (circ2);
\coordinate (circ1) at ($ (30:0.5) + (2.85,0.75) $);
\coordinate (circ2) at ($ (30:0.5) + (5.85,0.75) $);
\coordinate (circ3) at ($ (330:0.5) + (2.85,0.75) $);
\coordinate (circ4) at ($ (330:0.5) + (5.85,0.75) $);
\filldraw
	[
	fill=blue!22, 
	decoration={markings, mark=at position 0.23 with {\arrow{>}},
					mark=at position 0.745 with {\arrow{>}}
					}, postaction={decorate}
	] 
	(circ2) -- (circ1) arc (30:330:0.5) -- (circ3) -- (circ4) -- (circ2);
\filldraw
	[
	fill=blue!22, 
	decoration={markings, mark=at position 0.03 with {\arrow{>}},
					mark=at position 0.13 with {\arrow{>}},
					mark=at position 0.31 with {\arrow{>}},
					mark=at position 0.40 with {\arrow{>}},
					mark=at position 0.46 with {\arrow{>}},
					mark=at position 0.534 with {\arrow{>}},
					mark=at position 0.63 with {\arrow{>}},
					mark=at position 0.81 with {\arrow{>}},
					mark=at position 0.906 with {\arrow{>}},
					mark=at position 0.9635 with {\arrow{>}}
					}, postaction={decorate}
	] 
	(-0.25,3.2) -- (-0.25,1.25) -- (-6,1.25) -- (-6,0.75) -- (-0.25,0.75) -- (-0.25,-3.2) -- (0.25,-3.2) -- (0.25,-1.25) -- (6,-1.25) -- (6,-0.75) -- (0.25,-0.75) -- (0.25,3.2) -- (-0.25,3.2);
%
\fill (-2.85,-0.75) circle (3.3pt) node[left] {};
\fill (2.85,0.75) circle (3.3pt) node[right] {};
%
\draw (0,0) [black] node {};
\draw (0,-2.25) [black] node {};
\end{tikzpicture}
\right\rangle
=
\left\langle
\begin{tikzpicture}[very thick,scale=0.3,color=green!50!black, baseline, rounded corners=0.1pt]
\clip (0,0) ellipse (6cm and 3cm);
\nicedashedpalecolourscheme (0,0) ellipse (6cm and 3cm);
\draw
	[
	decoration={markings, mark=at position 0.45 with {\arrow{>}}}, postaction={decorate}
	] 
	(-2.85,-0.75) -- (-5.85,-0.75);
\draw
	[
	decoration={markings, mark=at position 0.45 with {\arrow{>}}}, postaction={decorate}
	] 
	(2.85,0.75) -- (5.85,0.75);
\draw
	[
	decoration={markings, mark=at position 0.2 with {\arrow{>}},
					mark=at position 0.5 with {\arrow{>}},
					mark=at position 0.84 with {\arrow{>}}
					}, postaction={decorate}
	] 
	(0,-3.2) -- (0,3.2);
\draw
	[
	decoration={markings, mark=at position 0.5 with {\arrow{<}}}, postaction={decorate}
	] 
	(0,-1) -- (6,-1);
\draw
	[
	decoration={markings, mark=at position 0.5 with {\arrow{>}}}, postaction={decorate}
	] 
	(0,1) -- (-6,1);
%
\fill (-2.85,-0.75) circle (4.3pt) node[below] {};
\fill (2.85,0.75) circle (4.3pt) node[above] {};
\fill (0,1) circle (4.3pt) node[below] {};
\fill (0,-1) circle (4.3pt) node[below] {};
%
\draw (0.9,-2) [black] node {{\scriptsize $T$}};
\draw (1.05,0) node {{\scriptsize $A_X$}};
\end{tikzpicture}
\right\rangle ,
\end{align*}
expressing a $T'$-correlator in terms of a $T$-correlator with a network of $A_X$-defects. 
As a special case~\eqref{eq:genorbequ} implies that boundary conditions of theory~$T'$ are in one-to-one correspondence with $A_X$-modules. 

Naturally the above constructions can be applied to Landau-Ginzburg models. If~$G$ is a finite symmetry group of a potential~$W$, then one finds~\cite[Thm.\,7.2]{genorb} that $G$-equivariant matrix factorisations of~$W$ are equivalent to modules over $A_G = \bigoplus_{g\in G}(I_W)_g$ where $(-)_g$ denotes twisting by~$g$ from the right. This recasts conventional Landau-Ginzburg orbifolds (including discrete torsion) in defect language, as worked out in detail in~\cite{BCP}. 

To construct other examples of generalised orbifolds between Landau-Ginzburg models with potentials~$W$ and~$V$, it suffices to find a defect $X:W\rightarrow V$ with invertible quantum dimension. In this case, as an immediate consequence of \eqref{eq:genorbequ}, we have an equivalence between matrix factorisations of~$V$ and $A_X$-modules.

To some degree equivalences of this form come very cheaply: once a candidate~$X$ is identified, our explicit residue expression~\eqref{eq:defectaction} makes computing $\dim(X)$ a straightforward exercise. While finding defects with invertible quantum dimension is still nontrivial, we nonetheless expect that many new equivalences can be produced in this way. For example, (generalised) orbifolds between A- and D-type minimal models were constructed in~\cite[Sect.\,7.3]{genorb}, and the CFT/LG correspondence predicts generalised orbifolds relating them to exceptional minimal models as well. We further expect generalised orbifolds involving Calabi-Yau compactifications \`{a} la~\cite{o0503632, hhp0803.2045}, and eventually for them to play a role in a generalisation of homological mirror symmetry to the defect sector. 

\medskip

\noindent\textbf{Acknowledgements. } 
We thank Ilka Brunner, Daniel Plencner and Ingo Runkel for helpful comments on the manuscript. 
Nils Carqueville is grateful to Marc Lehmacher and Natina Zulficar.

\appendix

\section{Zorro proof for beginners}\label{app:ZorroDirac}

For a general defect $X: W \longrightarrow V$ the proof of the Zorro moves in \cite{cm1208.1481} requires the introduction and careful study of Atiyah classes. Nevertheless the key idea is a simple one, and in this section we explain a special case where the proof reduces to a straightforward calculation with Pauli matrices $\sigma_1,\sigma_2,\sigma_3$. 

We take $W = 0$, $V = z_1^2 + z_2^2$ and 
\[
d_X = \begin{pmatrix} 0 & z_1 - \I z_2 \\ z_1 + \I z_2 & 0 \end{pmatrix} = \sigma_1 z_1 + \sigma_2 z_2\,.
\]
Then $X$ is a defect $W \rightarrow V$ and
\begin{align*}
\partial^{z,z'}_{[1]} d_X = \partial_{z_1} d_X  = \sigma_1 = \begin{pmatrix} 0 & 1 \\ 1 & 0 \end{pmatrix} , 
\quad 
\partial^{z,z'}_{[2]} d_X = \partial_{z_2} d_X  = \sigma_2 = \begin{pmatrix} 0 & -\I \\ \I & 0 \end{pmatrix}.
\end{align*}
Let $e_0, e_1$ constitute a $\C[z_1,z_2]$-basis of~$X$ with $|e_i| = i$ and note that $\sigma_3 = -\I \sigma_1 \sigma_2$ is the grading operator $\sigma_3(v) = (-1)^{|v|} v$. There are four Zorro moves to prove: the two in~\eqref{eq:Zorro} are trivial, so we focus our attention on the first identity in \eqref{eq:Zorro2} (the second is similar). Let $\mathcal{Z}$ denote the composite
\[
\xymatrix@C+2pc{
X \ar[r]^-{\lambda^{-1}} & I_V \otimes_{\C[z]} X \ar[r]^-{\coev \otimes 1} & X \otimes_{\C} X^\vee \otimes_{\C[z]} X \ar[r]^-{\rho\circ(1 \otimes \eval)} & X \,.
}
\]
We prove that this is the identity map. In this case our explicit formulas give
\begin{align*}
\lambda^{-1}(e_i) &= \sum_{l \geqslant 0} \sum_{a_1 < \cdots < a_l} \sum_j \theta_{a_1} \ldots \theta_{a_l} \{ \sigma_{a_l} \ldots \sigma_{a_1} \}_{ji} \otimes e_j\,, \\
\eval( \nu \otimes \eta ) &= \frac{\I}{4}(-1)^{|\eta|} \nu(\eta) \big|_{z_1=z_2=0} \, , \\
\coev( \theta_{a_1} \ldots \theta_{a_l} ) &= \sum_{i,j} (-1)^{\binom{r+1}{2} + s} \{ \sigma_{b_1} \ldots \sigma_{b_r} \}_{ij} \, e_i \otimes e_j^*
\end{align*}
where $\theta_{a_1} \ldots \theta_{a_l} \theta_{b_1} \ldots \theta_{b_r} = (-1)^s \theta_1 \theta_2$. Substituting, we find
\begin{align*}
\mathcal{Z}(e_i) &= (1 \otimes \eval) \sum_{l \geqslant  0} \sum_{a_1 < \cdots < a_l} \sum_j \coev( \theta_{a_1} \ldots \theta_{a_l} ) \{ \sigma_{a_l} \ldots \sigma_{a_1} \}_{ji} \otimes e_j\\
&= \sum_{l \geqslant 0} \sum_{a_1 < \cdots < a_l} \sum_{j,i',j'} (-1)^{\binom{r+1}{2} + s} \{ \sigma_{a_l} \ldots \sigma_{a_1} \}_{ji} \{ \sigma_{b_1} \ldots \sigma_{b_r} \}_{i'j'} e_{i'} \otimes \eval(e_{j'}^* \otimes e_j)\\
&= \sum_{l \geqslant 0} \sum_{a_1 < \cdots < a_l} \sum_{i'} \frac{\I}{4} (-1)^{\binom{r+1}{2} + s + |e_i| + l} \{ \sigma_{b_1} \ldots \sigma_{b_r} \sigma_{a_l} \ldots \sigma_{a_1} \}_{i'i} \, e_{i'}\\
&= \sum_{l \geqslant 0} \sum_{a_1 < \cdots < a_l} \sum_{i'} \frac{\I}{4} (-1)^{\binom{r+1}{2} + s + |e_i| + l + rl + \binom{l}{2}} \{ \sigma_{a_1} \ldots \sigma_{a_l} \sigma_{b_1} \ldots \sigma_{b_r} \}_{i'i} \, e_{i'}\\
&= \sum_{l \geqslant 0} \sum_{a_1 < \cdots < a_l} \sum_{i'} \frac{\I}{4} (-1)^{1 + |e_i|} \{ \sigma_1 \sigma_2 \}_{i'i} \, e_{i'} = \sum_{l \geqslant 0} \sum_{a_1 < \cdots < a_l} \frac{1}{4} \, e_i = e_i\,.
\end{align*}
In the last line we sum over the $2^m = 4$ basis elements of $I_V$, and this factor of~$4$ cancels with the~$\frac{1}{4}$ coming from the residue denominator in the evaluation map. In the general case something analogous happens: the quantum dimension of $I_V$ `cancels' with the residue, and the subtle part of the proof lies in showing that $\lambda^{-1}$ and the coevaluation map combine to produce the quantum dimension.

\end{document}